\documentclass[letterpaper,10pt]{article}
\usepackage{textcomp}
\usepackage{graphicx}
\usepackage{fancyhdr}
\usepackage{epsfig}

\title{\textbf{Is cosmic expansion of the universe accelerating?}}
\author{D. C. Choudhury\\ \textit{\small{Department of Physics, Polytechnic University, Brooklyn, New York 11201, USA}}\\ \textit{\small{e-mail: dchoudhu@duke.poly.edu}}}
\date{}

\def\abstract{\centering$\!$}

\begin{document}
\maketitle
\thispagestyle{empty}
\begin{abstract}
\parbox[s]{10cm}{\small\bf Abstract. \rm Currently available Type Ia distant supernovae observed data seem to support evidence that the cosmic expansion of the universe is accelerating. This unexpected result is beyond any standard model of modern cosmology. The new concept advanced to account for the acceleration is dark energy or quintessence with negative pressure. Most analyses using this new form of energy describe the observed data with great accuracy, although there has been no laboratory confirmation of it. The present work analyzes the consequences of Thomson scattering on Type Ia supernovae data for two significant reasons; (i) recently observed data reveal the existence of sufficient amount of ionized baryonic (hydrogen) dark matter in the intergalactic medium, a necessary ingredient for Thomson scattering, and (ii) its effects have not been considered previously in determining distances to the supernovae from their observed distance moduli. Quantitative results of the present investigation based on observed data and corrected for Thomson scattering are consistent with the prediction of Hubble expansion of the universe.}
\end{abstract}

\begin{keyword}
\parbox[l]{10cm}{\begin{tabular}{lp{0.7\linewidth}}\it Subject headings: \rm & cosmology:  theory - distances and redshifts - supernovae, Hubble expansion:  constituents of universe - missing dark matter  
\end{tabular}}
\end{keyword}

Measurements of distances to Type Ia distant supernovae (SNe Ia) appear to suggest expansion of the universe is accelerating~[1-4]. Precise determination of the first and second peaks in the anisotropy spectrum of cosmic microwave background radiation (CMBR) indicates that geometry of the universe is spatially flat, $\Omega_{tot}\cong 1$~[5-7]. No theoretical framework within the fundamental laws of physics exists that specifically accounts for the accelerating expansion. The most successful mechanism invoked is Einstein's cosmological constant $\Lambda$ in his 1917 equations which he later called "my greatest blunder", after Hubble first established in 1929 that the universe is expanding. But its interpretation as dark energy or quintessence with negative pressure [8-13], equivalent to gravitationally repulsive matter, is questionable since its calculated value is found to be about 120 orders of magnitude greater than the observed value for $\Lambda$~[14]. Also, dark energy has never been directly detected. Nevertheless, presently it is the only widely accepted explanation for accelerated expansion. The implications of cosmic acceleration are so profound for cosmology and our understanding of fundamental physics that it has been a subject of intense scrutiny and will continue to be so.

The present investigation examines whether the interpretation of the SNe Ia data [2,4] which concludes that expansion of the universe is accelerating, is valid or can it be accounted for within the framework of the known laws of physics, without invoking undetected dark energy. Thus an astrophysical interpretation of the observed data is searched for within the confines of the fundamental laws of physics. Since the effect of Thomson scattering, TS [15] in determining distances to SNe Ia has not been taken into account, it is investigated here for the following reasons: It is pure elastic, its total cross-section is constant, and it is independent of incident frequency. Thus it preserves the characteristics of SNe Ia spectra observed, relevant in the present analysis. Furthermore, recently published observational data [16-21] reveal the existence of a sufficient amount of dark matter consisting mostly of ionized hydrogen gas in warm-hot intergalactic medium (WHIM), an essential ingredient for TS. Calculations are performed within the framework of Friedmann-Robertson-Walker (FRW) cosmology [22] of Einstein's equations without cosmological constant for large-scale structure of the universe, for a special case of the flat universe. Consequently it is consistent with observations [5-7];  and compatible  with the inflationary model [23-25] of cosmology.

It is essential to search for the constituents and their respective fractional density of the universe because they are among the most important basic knowledge necessary for our understanding of modern cosmology.  The latest investigations [5-7,12-13] have led to the conclusions that the constituents and their respective fractional density in the unit of critical density are: baryons $\Omega_B=0.044 \pm 0.004$, non-baryonic dark matter $\Omega_{DM}=0.23 \pm 0.04$, and dark energy $\Omega_\Lambda=0.73 \pm 0.04$.

In light of the above [5-7, 12-13], the total matter density of the universe in units of the critical density $\Omega_{tot}\cong 1$. The observed visible baryonic matter density $\Omega_{BM}=0.044 \pm 0.004$. The   invisible non-baryonic dark matter $\Omega_{DM}=0.23 \pm 0.04$ is inferred from its gravitational effect on visible baryonic objects. Since this dark matter under the influence of gravity exhibits all characteristics of normal baryonic matter, most probably it is baryonic dark matter $\Omega_{BDM}$ present in the form of ionized intergalactic hydrogen plasma [26].  This inference is also supported by recently published observational data [16-21]. The remaining $\sim$73\% of dark matter has never been observed and very little is known about its nature. Hence in the present work, it is denoted as exotic dark matter $\Omega_{XDM}$. Discovering of this exotic dark matter is among the most challenging unresolved scientific problems in present day science.

Therefore within the framework of standard model of cosmology prior to late 1990's, the universe is considered to be homogeneous, isotropic and consisting only of ordinary visible baryonic matter $\Omega_{BM}$, dark baryonic matter $\Omega_{BDM}$, and exotic dark matter $\Omega_{XDM}$. Hence
\begin{equation}\label{eq:1}
\Omega_{TOT}\cong\Omega_{BM}+\Omega_{BDM}+\Omega_{XDM},
\end{equation}
where in the unit of critical density, $\Omega_{BM}=0.044 \pm 0.004$, $\Omega_{BDM}=0.23 \pm 0.04$, and $\Omega_{XDM}=0.73 \pm 0.04$. It is pertinent to point out here that magnitudes of these three cosmological parameters which enter into present investigation are equal to those of the cosmological parameters $\Omega_B$, $\Omega_{DM}$, and $\Omega_\Lambda$ respectively obtained from observed data [7,12] and summarized in page 16 of reference [13]. However there are significant differences in their microscopic properties as discussed above. Except for $\Omega_B$ and $\Omega_{BM}$, they are identical.

Since the method of application of TS in cosmology is given in reference [27], only main features of the formulation which are of importance in the present investigation are given below. The decrease in intensity of the radiation due to TS depends on the free electrons in the path from the source to the observer. Now consider radiation of intensity $I_S$ emitted by a Type Ia supernova located at redshift $z_S$ and received by an observer at redshift $z$. The fractional reduction of intensity $I(z)$ in traversing an infinitesimally small length of path $cdT_H(z)$ at redshift $z$ is
\begin{equation}\label{eq:2}
\frac{dI}{I(z)}=-\sigma_TN_e(z)cdT_H(z),
\end{equation}
where $\sigma_T$ is the total TS cross-section, $N_e(z)$ is the number of electron density, $c$ is the velocity of light, and $dT_H(z)$ is an infinitesimal interval of Hubble time at redshift $z$. After substituting the values of $N_e(z)$ and $dT_H(z)$ in eq. \ref{eq:2}, one obtains
\begin{equation}\label{eq:3}
\frac{dI}{I(z)}=\frac{9c}{16\pi G}\times \frac{\sigma_T\Omega_{BDM}H_0}{m_H}\times (1+z)^{1/2}dz,
\end{equation}
where $G$ is the gravitational constant, $m_H$ is the hydrogen mass, $H_0$ is the present value of the Hubble constant, and $\Omega_{BDM}$ is defined in eq. \ref{eq:1}.

Integrating over a path from the SNe Ia at redshift $z_S$ to an observer at redshift 
$z = 0$, gives
\begin{equation}\label{eq:4}
\frac{I_0}{I_S}=\exp\{-\Omega_{BDM}\frac{3c\sigma_TH_0}{8\pi Gm_H}[(1+z)^{3/2}-1]\},
\end{equation}
where redshift $z_S$ has been replaced by $z$. This reduction factor $I_0/I_S$ represents the loss of light intensity due to Thomson scattering in traversing the path from the source SNe Ia at $z = z_S$ to the point of observation at $z = 0$.

Observational data on redshift $z$ in the range of $0.0043 \leq z \leq 1.7$ and their corresponding  distance  modulus $m-M$ of SNe Ia, are given in references [2,4].  Data at $z$ = 0.43 and 0.48 are the mean values of two observations at each of these redshifts. These results are in striking agreement with those of the other competing groups such as in reference [3]. Observed luminosity distance $R_{OBS}$ in unit of Mpc is calculated using eq. \ref{eq:5} below, reference [28], for all available distance moduli.
\begin{equation}\label{eq:5}
R_{OBS}=10^{(m-M-25)/5}
\end{equation}

Theoretical luminosity distance $D_L$  in terms of redshift $z$, present value of Hubble constant $H_0$, and deceleration parameter $q_0$, within the framework of standard model of cosmology [26], p. 485 is given as,
\begin{equation}\label{eq:6}
D_L=\frac{c}{H_0q_0^2}[q_0z+(q_0-1)(\sqrt{1+2q_0z}-1)].
\end{equation}

The values of theoretical  luminosity distance $D_L$ for  all observed redshift $z$ are calculated using eq. \ref{eq:6}, for a flat universe [5-7], taking $q_0 = 1/2$, and currently accepted value for Hubble constant $H_0 = 71$ Km-s$^{-1}$-Mpc$^{-1}$ [7].

The correction factors, $I_0/I_S$ due to TS for $\Omega_{BDM}$=0.19, 0.23, and 0.27 are calculated from eq. \ref{eq:4}. The corrected distance moduli $(m - M)^\prime$ due to TS are obtained by multiplying observed modulus $(m - M)$ by $I_0/I_S$. Finally the corrected luminosity distances $D^\prime_L$ as a consequence of TS are computed using  $(m - M)^\prime$ from eq. \ref{eq:5} for $\Omega_{BDM}$ = 0.19, 0.23, and 0.27. The luminosity distances versus redshifts are plotted in 
Figures 1~(a), 1~(b), and 1~(c) 
for observed luminosity distances, Hubble luminosity distances, and corrected luminosity distances, data, corresponding to $\Omega_{BDM}$=0.19, 0.23, and 0.27 respectively. As can be seen from Fig. 1~(b),
 the best agreement with the theoretical Hubble luminosity distances (viz. solid  Hubble line), for all values of redshift $z$, from  0.0043 to 1.7, are in  good agreement with  those of the TS corrected luminosity distances corresponding to $\Omega_{BDM} = 0.23$ as depicted by diamond shapes in Fig. 1 (b). 
The justification for the present work has already been given in detail in the text.

In conclusion, results of the present investigation show that the currently available Type Ia supernovae observed data  [2-4], when corrected for TS, do not support the evidence for accelerating expansion of the universe. Furthermore, present investigation preserves the Einstein's theory of relativity intact; and   reveals that the 23\% of unknown invisible dark matter component of the universe ($\Omega_{DM}\cong 0.23$), inferred from its gravitational effect, is baryonic dark matter $\Omega_{BDM}$ present in the form of ionized intergalactic hydrogen plasma.

\clearpage
\section*{Appendix A: Figures}
\begin{figure}[h!]
\centering
\includegraphics[height=11.5cm]{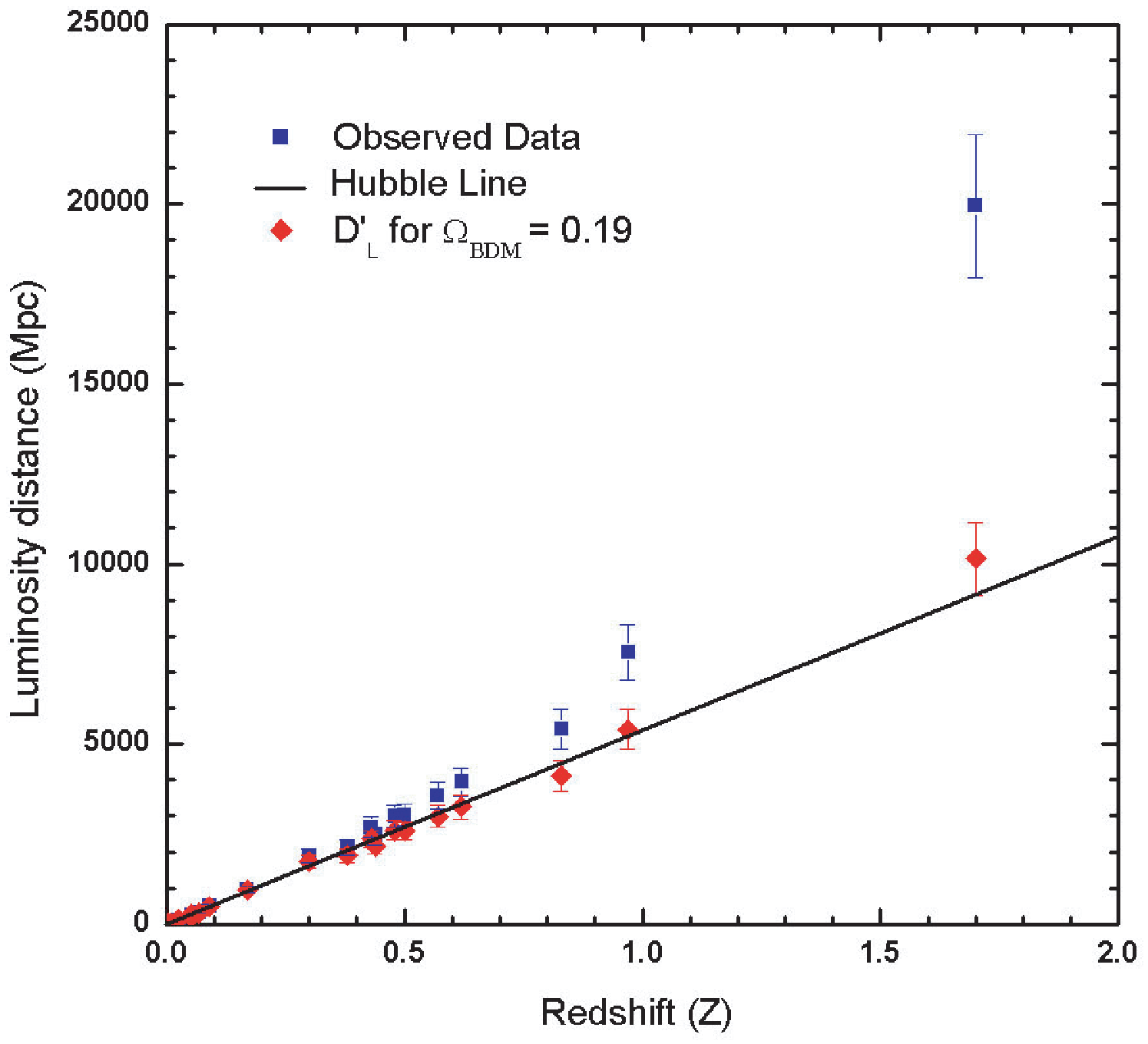} 
\label{fig:1a}
\vspace{-2.5cm}
\includegraphics[height=11.5cm]{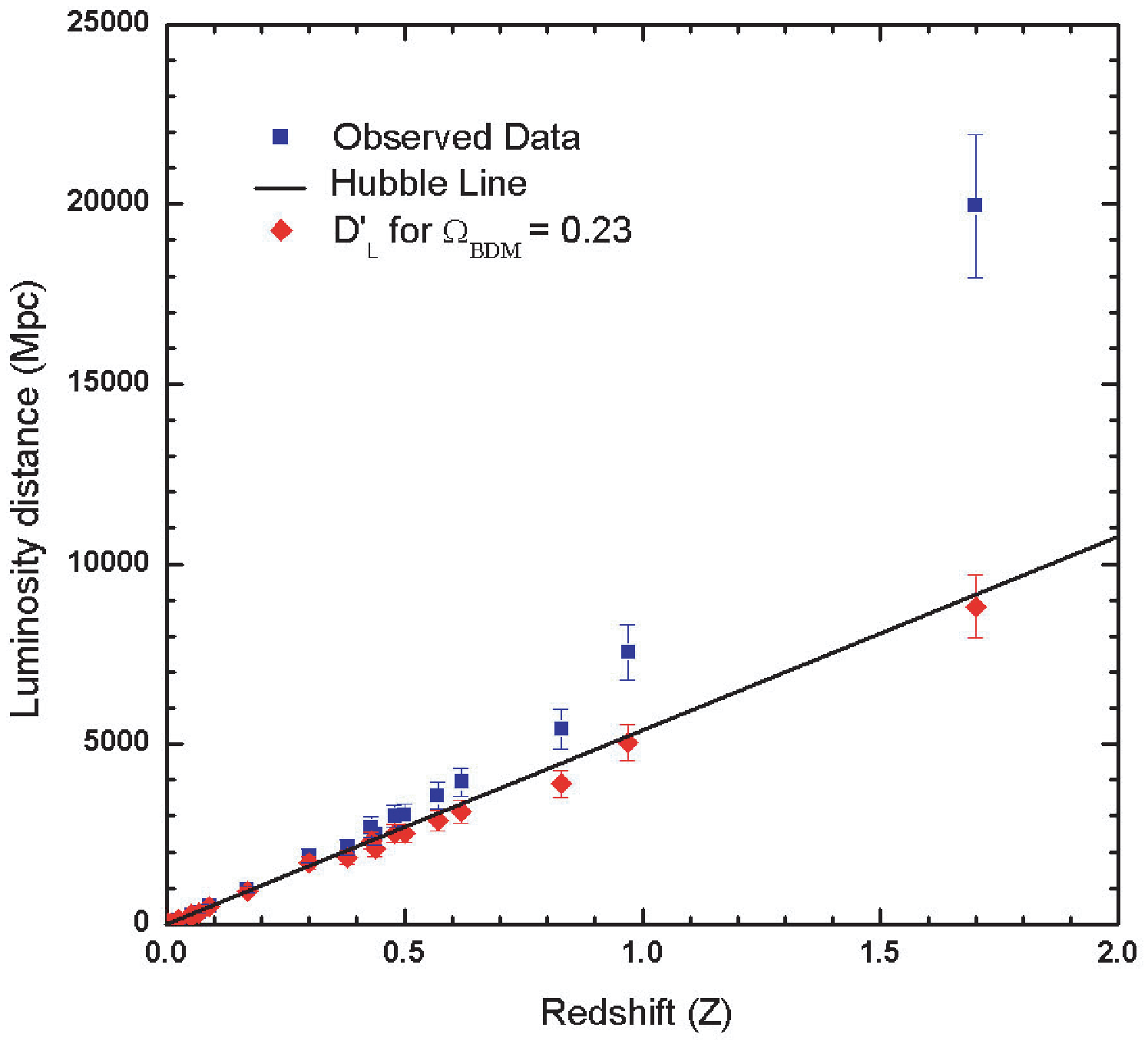} 
\label{fig:1b}
\vspace{-4cm}
\end{figure}
\vspace{-14cm}
\hspace{3cm}
Fig. 1(a)

\vspace{9cm}
\hspace{3cm}
Fig. 1(b)
\clearpage
\begin{figure}[th!]
\centering
\vspace{2cm}
\includegraphics[height=11.5cm]{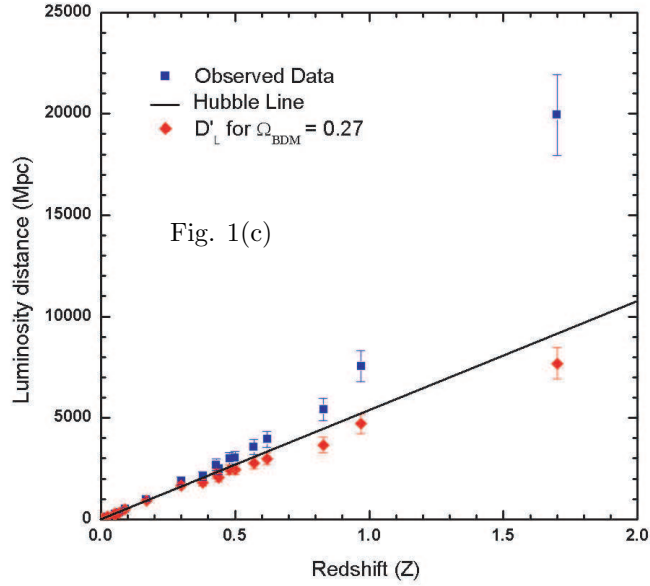} 
\caption{Hubble diagram of Type Ia supernovae data for a flat universe with Hubble constant $H_0 = 71$ Km-s$^{-1}$-Mpc$^{-1}$ [7]. The solid line represents theoretical Hubble luminosity distance $D_L$ calculated from eq. \ref{eq:6} versus the observed red-shift $z$ given in references [2,4]. The square shapes represent the values of observed luminosity distance $R_{OBS}$ obtained using distance moduli given in references [2,4] versus red-shift $z$. The diamond shapes represent the Thomson scattering-corrected luminosity distances $D^\prime_L$ obtained in present work for $\Omega_{BDM}$ = 0.19 -- fig. 1(a), $\Omega_{BDM}$ = 0.23 -- fig. 1(b), and $\Omega_{BDM}$ = 0.27 -- fig. 1(c), respectively. Fig. 1(b), for $\Omega_{BDM}$ = 0.23, shows that the observed luminosity distances $R_{OBS}$ when corrected for TS are in good agreement with the prediction of the Hubble law  expansion without cosmic acceleration.}
\label{fig:1c}
\end{figure}
\vspace{-13cm}
\hspace{3cm}
Fig. 1(c)
\clearpage

\thispagestyle{empty}
\begin{figure}[t]
\vspace{-4cm}
\hspace{-3cm}
\includegraphics[angle=180,width=1.5\textwidth]{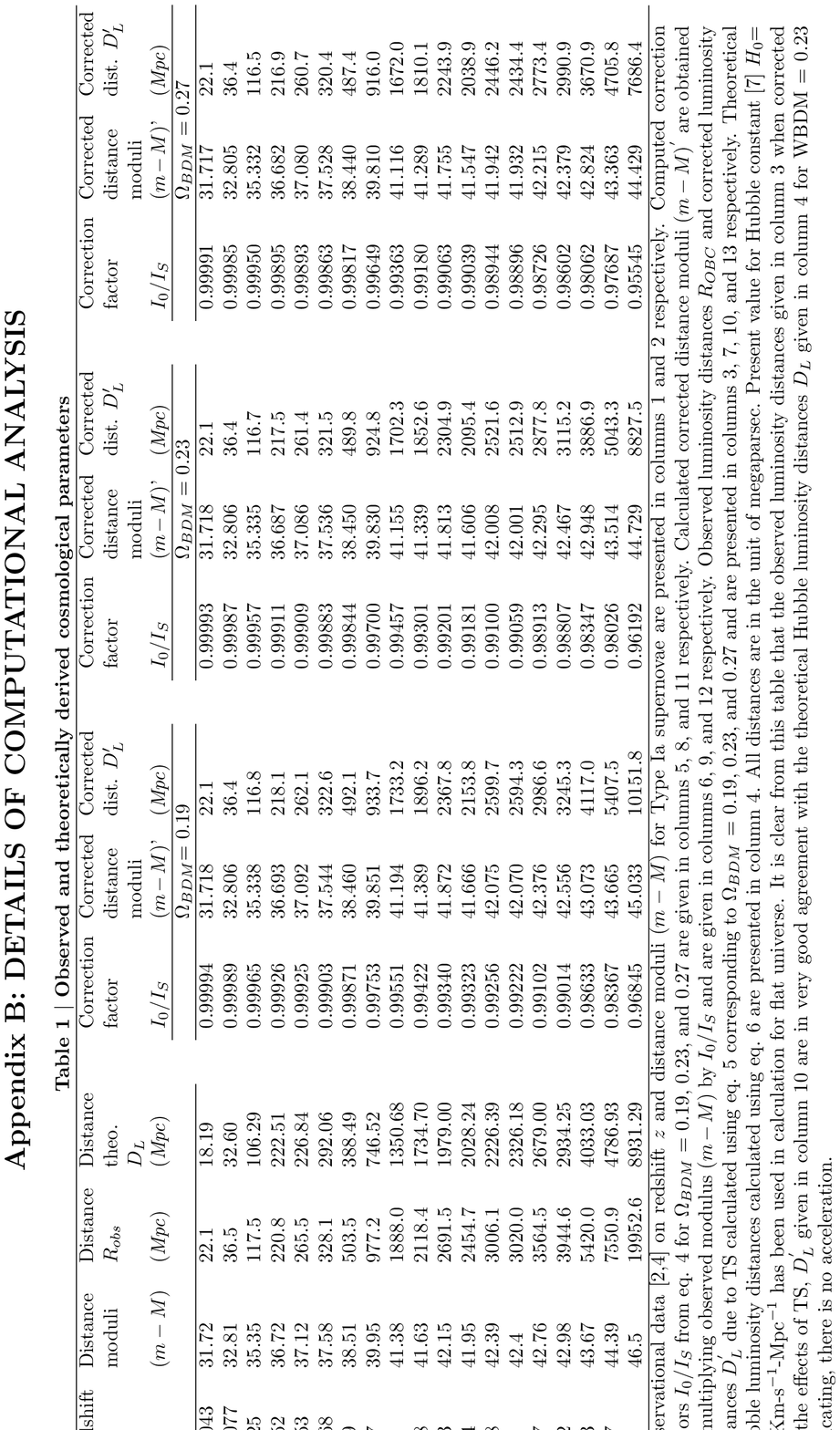}
\end{figure}

\end{document}